\documentclass[aps,reprint,twocolumn,preprintnumbers,floatfix,nofootinbib]{revtex4-1}

\usepackage[T1]{fontenc} 
\usepackage[utf8]{inputenc}

\usepackage[dvipdfm]{graphicx}
\usepackage{mathrsfs}
\usepackage{amssymb}
\usepackage{verbatim}
\usepackage{color}
\usepackage[dvipsnames]{xcolor}
\usepackage{multirow}
\usepackage{amsmath}
\usepackage{slashed} 
\usepackage{float}
\usepackage[normalem]{ulem}
\usepackage{sidecap}
\usepackage{hyperref}
\usepackage{cancel}
\usepackage{enumerate}
\hypersetup{pdfstartview=FitV,colorlinks=true,linkcolor=blue,citecolor=red,filecolor=black,urlcolor=blue}

\newcommand{\beq}{\begin{equation}}
\newcommand{\eeq}{\end{equation}}
\newcommand{\bea}{\begin{eqnarray}}
\newcommand{\eea}{\end{eqnarray}}

\def\trh{T_{\rm RH}}

\usepackage[parfill]{parskip}
\begin{document}

 \preprint{UMN--TH--4511/25, FTPI--MINN--25/13,
 KCL--PH--TH/2025--39, CERN--TH--2025--198}

\vspace*{1mm}

\title{Effects of Radiative Corrections on Starobinsky Inflation}

\author{John Ellis$^{a,b}$}
\email{John.Ellis@cern.ch}
\author{Tony Gherghetta$^{c}$}
\email{tgher@umn.edu}
\author{Kunio Kaneta$^{d}$}
\email{kaneta@ed.niigata-u.ac.jp}
\author{Wenqi Ke$^{c,e}$}
\email{wke@umn.edu}
\author{Keith A. Olive$^{c,e}$ \vspace{0.3cm}
}
\email{olive@umn.edu}
\vspace{0.5cm}

\affiliation{${}^a $Department of Physics, King’s College London, Strand, London, WC2R 2LS, UK}
\affiliation{${}^b $Theoretical Physics Department, CERN, Geneva, Switzerland}
\affiliation{${}^c$School of
 Physics and Astronomy, University of Minnesota, Minneapolis, MN 55455,
 USA}
\affiliation{${}^d $Faculty of Education, Niigata University, Niigata 950-2181, Japan}
\affiliation{${}^e$William I. Fine Theoretical Physics Institute, School of
 Physics and Astronomy, University of Minnesota, Minneapolis, MN 55455,
 USA}

\date{\today}

\begin{abstract}
We analyze radiative corrections to the Starobinsky model of inflation arising from self-interactions of the inflaton, and from its Yukawa couplings, $y$, to matter fermions, and dimensionful trilinear couplings, $\kappa$, to scalar fields, which could be responsible for reheating the Universe after inflation. The inflaton self-interactions are found to be of higher order in the Hubble expansion rate during inflation, and hence unimportant for CMB observations. In contrast, matter couplings to the Starobinsky inflaton can have significant effects on the spectral index of scalar CMB perturbations, $n_s$, and on the tensor-to-scalar ratio, $r$. Using a renormalization-group improved analysis of the effective inflationary potential, we find that the \textit{Planck} measurement of $n_s$ constrains the inflaton coupling to light fermions in the Einstein frame: $y < 4.5 \times 10^{-4}$, corresponding to an upper limit on the reheating temperature $T_{\rm RH} < 2 \times 10^{11}$~GeV, whereas the ACT DR6 measurement of $n_s$ corresponds to $3.8 \times 10^{-4} < y < 5.6 \times 10^{-4}$ and $1.7 \times 10^{11} ~{\rm GeV} < T_{\rm RH} < 2.8 \times 10^{11}$~GeV, while the upper limits on $r$ provide weaker constraints. \textit{Planck} data also imply a constraint on a trilinear inflaton coupling to light scalars in the Einstein frame: $\kappa \leq 4 \times 10^{12}$~GeV, corresponding to $T_{\rm RH} \leq 4.2 \times 10^{13}$~GeV. We further present constraints on inflaton couplings to massive fermions and scalars, and analyze constraints on couplings in the Jordan frame.
\end{abstract}

\maketitle

\section{Introduction}

 Until recently, data on perturbations \cite{Planck} in the cosmic microwave background (CMB) have been strikingly consistent with the original model of inflation formulated by Starobinsky \cite{Staro} as an $R^2$ deformation of the minimal Einstein-Hilbert action, which could be reformulated \cite{WhittStelle,Kalara:1990ar}, as the minimal action supplemented by a scalar field with a characteristic tree-level potential 
\begin{equation}
\label{treestaro}
V \; = \; \frac34 m^2 M_P^2\left[ 1 - e^{-\sqrt{\frac23} \frac{\phi}{M_P} }\right]^2 \, ,
\end{equation}
where the reduced Planck mass $M_P = \sqrt{8\pi/G_N} = 2.435 \times 10^{18}$~GeV.
From this potential, it is straightforward to compute the tilt of the scalar anisotropy spectrum, $n_s \simeq  0.965$, and a tensor-to-scalar ratio of $r \simeq 0.0035$, with a slight dependence on the reheating temperature after inflation. The {\it Planck} 2018 determination of $n_s$ is \cite{Planck}
\beq
n_s \; = \; 0.9649 \pm 0.0042 \; (68\%~{\rm C.L.}) \, ,
\label{nsexp}
\eeq 
and \textit{Planck} in combination with observations by BICEP/Keck~\cite{BICEP2021} has
provided an upper limit on $r$:
\beq
r \; < \; 0.036 \, ,
\label{rlim}
\eeq
which are both in excellent agreement with the predictions of the tree-level Starobinsky potential \eqref{treestaro}. 

However, this tree-level potential requires protection from higher-order corrections, particularly because its predictions for the CMB observables are very sensitive to the form of the potential at inflaton field values $\gtrsim 5 M_P$. In the original modified gravity formulation, the plateau part of the potential arises if the $R^2$ term dominates, but quantum gravity effects can generate $R^n$ ($n>2$) and other curvature invariants that distort the tree-level potential, so these must be tuned away.\footnote{Some constraints on higher curvature terms are discussed in, e.g., \cite{Copeland:2013vva} in the context of asymptotic safety.} In supergravity, the Starobinsky potential can arise using a no-scale K\"ahler potential $K$ and an appropriate superpotential $W$ \cite{eno6,eno7,enov1}, but higher-dimensional operators in $K$ or $W$ typically spoil the flatness unless their coefficients are tuned or forbidden by a symmetry. Moreover, in a UV extension of the inflationary model, couplings between the inflaton and heavy fields such as GUT-scale scalars can be generated, which tend to lift the potential at large field values and and perturb the predictions of the Starobinsky model \cite{deform}.

Once such higher-order terms and additional renormalizable couplings are suitably controlled or tuned, the next step is to consider loop corrections from fields that must couple to the inflaton to reheat the Universe after inflation. These scalar or fermion loops introduce genuine quantum effects that are distinct from the tree-level tuning of the potential and can be computed systematically within the effective field theory.

Inflaton decay is the most efficient mechanism for reheating the Universe after inflation~\cite{dg,afw,nos}. Indeed, if the matter Lagrangian is introduced in the Jordan frame (where the gravitational Lagrangian contains the $R^2$ correction),  there is a coupling between the inflaton and the SM Higgs boson, which leads to an efficient inflaton decay channel to a pair of Higgs bosons. Moreover, couplings between the inflaton and fermions may be introduced in either the Jordan or Einstein conformal frames. Whilst these couplings would not affect the tree-level potential, they would induce radiative corrections at the one-loop level, which may have observable effects on the predictions for the CMB observables. 

These radiative (quantum) corrections to the Starobinsky potential induced by the couplings related to inflaton decay are the main focus of this paper.  There have been previous studies  \cite{Kallosh:2016gqp,Kazakov:2023tii,Gialamas:2025kef,Wolf:2025ecy,Ahmed:2025rrg,Han:2025cwk} of their effects on the effective inflationary potential and on CMB observables, but calculations of such effects have gained increased relevance and urgency from the measurement of the scalar tilt 
reported recently by the Atacama Cosmology Telescope (ACT) collaboration \cite{ACT:2025fju}, which seems to deviate from the prediction of the Starobinsky model. Many possible explanations of this deviation have been proposed \cite{Kallosh:2025rni,Gialamas:2025kef,Dioguardi:2025vci,Antoniadis:2025pfa,Salvio:2025izr,German:2025mzg,He:2025bli,Drees:2025ngb,Zharov:2025evb,Haque:2025uri,Liu:2025qca,Gialamas:2025ofz,Byrnes:2025kit,Addazi:2025qra,Mondal:2025kur,Saini:2025jlc,Haque:2025uga,Hai:2025wvs,Heidarian:2025drk,Choudhury:2025vso,Pallis:2025gii,German:2025ide,deform,Aoki:2025ywt,Modak:2025bjv,Pallis:2025vxo}. Here we investigate systematically various sources of radiative corrections within the framework of the Starobinsky model, and use their consistency with CMB data to constrain possible couplings of the Starobinsky inflaton to other fields. Furthermore, because these couplings are directly related to the reheating temperature, we use constraints on these couplings to derive constraints on the reheating temperature.

In the following we first review in Section~\ref{sec:staro} the original Starobinsky model derived from the $R^2$ deformation of the minimal Einstein-Hilbert action. Then, as a warm-up exercise, in Section~\ref{1loop} we consider the 1-loop corrections to this theory due to inflaton self-interactions. We show that these are ${\cal O}(H^4)$ in general, where $H$ is the Hubble expansion rate during inflation, and hence have negligible impact on CMB observables.

However,  the picture changes in Section~\ref{1loopf}, where we consider the effects of radiative corrections due to the possible Yukawa coupling, $y$, of the Starobinsky inflaton to a matter fermion, such as may play a role in the reheating of the Universe following the end of inflation. We first compute the RGE-improved one-loop effective potential.
For small fermion masses, we set limits on $y$, and for large (GUT-scale) masses we set constraints on the fermion mass for a given value of $y$. 
We also discuss the corrections to the potential that arise when the coupling to fermions is introduced in the Jordan frame. In this case, we can set a limit on the fermion mass independent of the coupling $y$. 

We find that the value of $y$ is significantly constrained by measurements of $n_s$ and, to a lesser extent, by the upper limit on the magnitude of $r$.
The {\it Planck} measurement of $n_s$ constrains the coupling of the inflaton to a light fermion in the Einstein frame to $y < 4.5 \times 10^{-4}$, corresponding to an upper limit on the reheating temperature $T_{\rm RH} < 2 \times 10^{11}$~GeV, while the ACT DR6 measurement of $n_s$ corresponds to $3.8 \times 10^{-4} < y < 5.6 \times 10^{-4}$ and $1.7 \times 10^{11} ~{\rm GeV} < T_{\rm RH} < 2.8 \times 10^{11}$~GeV. We also present constraints on the inflaton coupling to fermions with a non-negligible bare mass $m_f$. For example, for $y=10^{-4}$, $m_f < 1.2 \times 10^{16}$~GeV,  while ACT DR6 requires $10^{16}~{\rm GeV} < m_f < 2\times 10^{16}~{\rm GeV}$. If the coupling is generated through a mass term in the Jordan frame, we estimate somewhat weaker bounds around $ 6 \times 10^{16}$~GeV. 

 In Section~\ref{1loopb} we introduce the possibility of inflaton decay to a scalar pair via a coupling $\kappa$, and consider the corrections to the Starobinsky potential that would be induced by such a coupling. We consider both couplings to light scalars such as the Standard Model Higgs boson and to possible GUT-scale scalars. As in the fermion case, the effects of these corrections on CMB observables provide limits on the coupling and on the reheating temperature when reheating is due to inflaton decay to these scalars. For light scalars, we find that $\kappa < 4 \times 10^{12}$~GeV, implying that $\trh < 4.2 \times 10^{13}$~GeV. However, the ACT DR6 value of $n_s$ is never reached. For a massive scalar the {\it Planck} 2018 bound is saturated at $m_s < 1.6 \times 10^{16}$~GeV for $\kappa = 10^{12}$~GeV.   We also derive limits on the scalar masses when the scalar Lagrangian is introduced in the Jordan frame. The possible role of supersymmetry and supersymmetry breaking is discussed in Section \ref{susy} and our conclusions are given in Section \ref{summary}.

\section{Starobinsky inflation}
\label{sec:staro}

Before considering radiative corrections, we first review the Starobinsky model of inflation~\cite{Staro}.
We denote the metric and the Ricci scalar in the Jordan frame as $\widetilde g_{\mu\nu}$ and $\widetilde R$, respectively.
The relevant action is given by
\begin{align}
    S &=
    \int d^4x\sqrt{-\widetilde g}\left[
        -\frac{M_P^2}{2}\widetilde{R} + \alpha \widetilde R^2 + {\cal L}_{\rm matter}
    \right] \, ,
\end{align}
where $\alpha$ is a dimensionless constant that will be related later to the inflaton mass, and we include a matter-sector Lagrangian ${\cal L}_{\rm matter}$. 
Introducing an auxiliary field $\Phi$ \cite{WhittStelle,Barrow:1988xh,Kalara:1990ar,eno9,building}, one can rewrite the action as
\begin{align}
    S &=
    \int d^4x \sqrt{-\widetilde g}\left[
        -\frac{M_P^2}{2} \widetilde{R} + 2 \alpha \Phi \widetilde{R} -\alpha \Phi^2
         + {\cal L}_{\rm matter}
    \right],
\end{align}
where the condition $\delta S/\delta \Phi=0$ imposes $\Phi=\widetilde R$.

To express the gravity sector in the canonical form of the Einstein-Hilbert action, we perform a conformal transformation to the Einstein frame
\begin{align}
 g_{\mu\nu}    &= e^{2\Omega}  \widetilde g_{\mu\nu}\, ,
 \label{gconf}
\end{align}
where $\Omega$ can be expressed as
\begin{align}
    e^{2\Omega} &=
    \left(
        1-\frac{4 \alpha\Phi}{M_P^2}
    \right) \, .
    \label{Oconf}
\end{align}
Using $e^{-2\Omega} \widetilde R=R-6g^{\mu\nu}\partial_\mu\ln\Omega \ \partial_\nu\ln\Omega$, we obtain
\begin{align}
    S &=
    \int d^4x\sqrt{-g}\left[
        -\frac{M_P^2}{2}R + 3M_P^2\ g^{\mu\nu}\ \partial_\mu \Omega\ \partial_\nu \Omega\right.\nonumber\\
        &\left. - \frac{M_P^4}{16\alpha}(1-e^{-2\Omega})^2 + e^{-4\Omega}{\cal L}_{\rm matter}
    \right] \, .
\end{align}
We can then define the canonical scalar field $\phi$ through
\begin{align}
    \Omega &= \frac{\phi}{\sqrt{6}M_P},
\end{align}
and the action becomes 
\begin{align}
\label{staraction}
    S &=
    \int d^4x\sqrt{-g}\left[
        -\frac{M_P^2}{2}R + \frac{1}{2} g^{\mu\nu}\partial_\mu\phi \ \partial_\nu\phi \right.\nonumber\\
        & \left.- \frac{3}{4}m^2M_P^2\left(
            1-e^{-\sqrt{\frac{2}{3}}\frac{\phi}{M_P}}
        \right)^2 + \Omega^4{\cal L}_{\rm matter}(\Omega)
    \right] \, ,
\end{align}
where $m$ is the inflaton mass obtained from the relation $\alpha=M_P^2/12m^2$.
We also see in (\ref{staraction}) the appearance of the Starobinsky potential (\ref{treestaro}) in the action in the Einstein frame. 

In general, one can relate the inflationary slow-roll parameters to derivatives of the potential:
\begin{equation}
\epsilon \; = \; \frac{M_P^2}{2} \left( \frac{V^\prime}{V} \right)^2 \; , \; \eta = M_P^2 \frac{V^{\prime \prime}}{V} \, ,
\end{equation}
at leading order, where a prime denotes a derivative with respect to the canonically-normalized inflaton field, $\phi$. The CMB observables can then be calculated in terms of these slow-roll parameters:
\begin{equation}
r \; = \; 16 \epsilon_* \; , \; n_s \; = \; 1 - 6 \epsilon_* +2 \eta_* \, ,\label{nseq}
\end{equation}
where the $_*$ subscript denotes a suitable pivot scale, often taken to be  $k_* = 0.05$~Mpc$^{-1}$.
The number of $e$-folds starting from
the horizon exit at the pivot scale to the end of inflation can also be computed in the slow-roll approximation:
\begin{equation}
N_* \;\equiv\; \ln\left(\frac{a_{\rm{end}}}{a_*}\right) \; = \; \int_{t_*}^{t_{\rm{end}}} H dt \; \simeq \;  - \int^{\phi_{\rm{end}}}_{\phi_*} \frac{1}{\sqrt{2 \epsilon_*}} \frac{d \phi}{M_P} \, ,
\label{e-folds}
\end{equation}
where $a_*$ is the value of the cosmological scale factor at the pivot scale, $a_{\rm end}$ is its value at the end of inflation (when exponential expansion ends and $\ddot{a}=0$) and $\phi_*$, and $\phi_{\rm end}$ are the corresponding values of $\phi$ 
at those two epochs. At the end of inflation, the slow-roll parameters also satisfy:
\begin{equation}
    \epsilon\simeq \left(1+\sqrt{1-\eta/2}\right)^2\,.
    \label{phiendeq}
\end{equation}
Assuming no additional entropy production after reheating, the number of $e$-folds is given by \cite{LiddleLeach,Martin:2010kz,EGNO5,egnov}:
\begin{equation}
\begin{aligned}
\label{eq:nstarreh}
N_{*} &= \\
& \hspace{-5mm} \ln \left[\frac{1}{\sqrt{3}}\left(\frac{\pi^{2}}{30}\right)^{1 / 4}\left(\frac{43}{11}\right)^{1 / 3} \frac{T_{0}}{H_{0}}\right]-\ln \left(\frac{k_{*}}{a_{0} H_{0}}\right) -\frac{1}{12} \ln g_{\mathrm{RH}}\\
&
\hspace{-5mm} +\frac{1}{4} \ln \left(\frac{V_{*}^{2}}{M_{P}^{4} \rho_{\mathrm{end}}}\right) +\frac{1-3 w_{\mathrm{int}}}{12\left(1+w_{\mathrm{int}}\right)} \ln \left(\frac{\rho_{\mathrm{R}}}{\rho_{\text {end }}}\right) 
\, ,
\end{aligned} 
\end{equation}
where $H_0$ is the present Hubble parameter, for which {\it Planck} found the best-fit value $H_0 = 67.36 \, \rm{km \, s^{-1} \, Mpc^{-1}}$~\cite{Planck}, and the present photon temperature $T_0 = 2.7255 \, \rm{K}$~\cite{Fixsen:2009ug}. The quantities $\rho_{\rm{end}} \equiv \rho(\phi_{\rm end})$ and $\rho_{\rm{R}}$ are, respectively, the energy density at the end of inflation and at the beginning of the radiation-dominated era when $w = p/\rho = 1/3$, $a_0 = 1$ is the present scale factor, $g_{\rm{RH}} =  427/4$ is the effective number of relativistic degrees of freedom in the SM during reheating, and $w_{\rm int}$ is the average equation-of-state parameter during reheating. For the choice
 $k_* \; = \; 0.05 \, \rm{Mpc}^{-1}$ of the pivot scale, the first line on the right-hand side of (\ref{eq:nstarreh}) gives
$N_*  \simeq  61.10$. The second line of (\ref{eq:nstarreh}) shows that
$N_*$ depends on the reheating temperature through $\rho_{\rm R}$. The inflationary observables, most notably $n_s$, also depend on the reheating temperature. 

The {\it Planck} 2018 determination of $n_s$ is given in Eq.~(\ref{nsexp}). Recently two ground-based experiments have released new results for $n_s$. ACT DR6 results \cite{ACT:2025fju} indicate a larger value of $n_s$. Their data when combined with {\it Planck}, lensing and DESI DR2 BAO results \cite{DESI:2025zpo} give 
\beq
n_s = 0.9752 \pm 0.0030 \; (68\%~{\rm C.L.}) \, .
\label{ACTns}
\eeq
In addition to the {\it Planck} result in Eq.~(\ref{nsexp}),
we will use this ACT result (\ref{ACTns}) to set limits on the 1-loop contributions to the inflaton potential. 
The South Pole Telescope (SPT) Collaboration has provided SPT-3G data \cite{SPT-3G:2025bzu} that, in contrast to ACT DR6, are in good agreement with {\it Planck}. The combination of SPT-Planck-ACT give $n_s = 0.9684 \pm 0.0030$ at 68\% CL.   

Finally, the parameter $m$ is determined from the amplitude of CMB density fluctuations, $A_s = 2.1 \times 10^{-9}$ \cite{Planck}. In the Starobinsky model (\ref{treestaro})  this amplitude is given by
\beq
A_s   =   \frac{V(\phi_*)}{24\pi^2 \epsilon_* M_P^4} =  \frac{3 m^2}{8\pi^2 M_P^2} \sinh^4 \left(\frac{\phi_*}{\sqrt{6}M_P} \right)\, , \label{As2}
\eeq
and fixes $m$ for any given value of $\phi_*$ that is determined by $\trh$.  For example, for $\trh \sim 4\times 10^{10}$~GeV, $\phi_* \simeq 5.3 M_P$ and
$m  \approx 1.3 \times 10^{-5}$ in Planck
units.

\section{One-Loop Corrections from Inflaton Self-Interactions}
\label{1loop}

The tree-level Starobinsky potential (\ref{treestaro}) appears explicitly in the action (\ref{staraction}). 
During inflation the exponential $e^{-\sqrt{2/3} \frac{\phi}{M_P} } \ll 1$, and we may
drop the terms in $V$ of higher order in the exponential, so that
\begin{equation}
\label{Vapprox}
V\simeq\frac34 m^2 M_P^2\left( 1 - 2 e^{-\sqrt{\frac23} \frac{\phi}{M_P}} +\dots \right) \, .
\end{equation}
We then have
\begin{equation}
\label{VprimeVdoubleprime}
V^\prime = \sqrt{\frac{3}{2}}m^2 M_P e^{-\sqrt{\frac23} \frac{\phi}{M_P}}, \quad V^{\prime \prime} = - m^2 e^{-\sqrt{\frac23} \frac{\phi}{M_P}}\,,
\end{equation}
in the approximation (\ref{Vapprox}).

In calculating the one-loop (Coleman-Weinberg) correction to the effective potential, it is important to note that during inflation the effective potential receives contributions from spacetime curvature. These can be derived using the heat kernel technique \cite{Markkanen:2018bfx}.~\footnote{In \cite{Markkanen:2018bfx}, the effective potential was computed assuming that the scalar field was a subdominant spectator, in which case the metric fluctuations can be neglected. When the inflaton dominates the energy density, gravity backreacts through the mixing of its scalar component. However, the resulting correction to the effective mass is of order the slow-roll parameter $\epsilon$, which is ${\cal O}(e^{-\sqrt{2/3} \frac{\phi}{M_P}})$, so one can neglect the effect of the metric fluctuations. This approximation is valid in the Einstein frame, as argued in~\cite{George:2013iia}.} Neglecting at this stage couplings of the Starobinsky inflaton to other fields, its effective mass-squared is
\begin{equation}
    \mathcal{M}_\phi^2= V ^{\prime\prime}(\phi)+ \frac{1}{6}R \simeq V^{\prime\prime}(\phi)-2H^2 \, ,
\end{equation}
where we have used $R\simeq -12H^2$ and $H^2\simeq V(\phi)/3M_P^2$ during slow-roll inflation. 

Choosing the $\overline{\text {MS}}$ renormalization scheme, the one-loop correction to the effective potential is \cite{Markkanen:2018bfx}: 
\begin{equation}
\Delta V=    \frac{1}{64\pi^2} \left[ \mathcal{M}_\phi ^4 \left(\log \frac{ |\mathcal{M}_\phi ^2| }{\mu^2}-\frac{3}{2}\right)+2 b_s \log \frac{|\mathcal{M}_\phi^2| }{\mu^2}\right] \, ,
\label{loopscalar}
\end{equation}
where $\mu$ is an arbitrary renormalization scale and the coefficient $b_s$ is given by
\begin{equation}
      b_s=-\frac{1}{180}R_{\mu\nu}R^{\mu\nu} +\frac{1}{180}R_{\mu\nu\rho\sigma}R^{\mu\nu\rho\sigma}\simeq -\frac{H^4}{15} \, .
\end{equation}
We see from (\ref{VprimeVdoubleprime}) that during inflation $V^{\prime\prime} \ll H^2$, so that $\mathcal{M}_\phi^2\simeq -2H^2 $. If we choose the renormalization scale $\mu$ so that the $\log$ is of order one,~\footnote{This is a generic choice, but the renormalization scale  may in principle be chosen to minimize $\Delta V$.} then the correction is of order $\Delta V\sim H^4 $.  Given that $3M_P^2 H^2=\rho_\phi\simeq V  $, we have $\Delta V /V \sim H^2/M_P^2\sim  10^{-10 }$. Therefore, the one-loop correction of the inflaton self-interaction is negligible. 

The one-loop potential \eqref{loopscalar} is derived in the small proper time (large-momentum) limit, so it captures  only the UV structure \cite{Markkanen:2018bfx}.
This can be seen, for example, from the form of the effective mass in \eqref{loopscalar}, which is taken to be its absolute value inside the logarithm. A negative effective mass would naively lead to imaginary terms
($\propto i\pi $), but we expect that if the effective potential correctly included the low momentum modes, these would regulate any apparent instability.

\section{One-Loop Corrections from Inflaton Couplings to Fermions}
\label{1loopf}

\subsection{Yukawa interaction with the inflaton}
\label{sec:FermionEinstein}

One may consider corrections to the inflaton potential due to a Yukawa coupling of a Dirac fermion to the inflaton of the form $y\bar{f}f\phi $, or $y\bar{f^c}f\phi/2 $ if $f$ is a Majorana fermion. In this section, we will focus on couplings to a Majorana fermion (e.g. this could be a right-handed neutrino). 
The effective, field-dependent fermion mass during inflation including the curvature contribution is
\begin{equation}
    \mathcal{M}^2_f=(m_f+y\phi)^2-\frac{1}{12}R
\simeq (m_f+y\phi)^2+H^2\,.
\end{equation} 
The Yukawa coupling may be solely responsible for reheating after inflation, in which case one can relate the coupling $y$ to the reheating temperature, $\trh$. The reheating temperature is obtained from equating the energy density in the inflaton with the energy density in radiation produced by inflaton decays, namely
\beq
g_{\rm RH} \frac{\pi^2}{30} \trh^4 = \frac{12}{25} (\Gamma_\phi M_P)^2 = \frac{3y^4}{1600\pi^2}  m^2 M_P^2 \,,
\label{eq:TRH}
\eeq
where $g_{\rm RH}$ is the number of relativistic degrees of freedom at $\trh$, $\Gamma_\phi$ is the inflaton decay rate and we have assumed a decay to a Majorana fermion in the second equality.

\subsubsection{Massless fermion $m_f=0$}
Initially, we consider a massless fermion coupled to the inflaton, and set $m_f = 0$.
The one-loop effective potential induced by the fermion coupling\footnote{Note that the one-loop effective potential also contains quadratic divergences and fermion tadpoles, which would shift the inflaton mass or vacuum. In the absence of supersymmetry, these contributions are assumed to be tuned to zero and we focus solely on the logarithmic terms shown in \eqref{yukawaloop}.} is~\cite{Markkanen:2018bfx}
\begin{equation}
    \Delta V _f =-\frac{n_i}{64\pi^2}\left[\mathcal{M}_f^4\left(\log \frac{|\mathcal{M}_f^2|}{\mu^2}-\frac{3}{2}\right)+\frac{1}{2} b_f \log \frac{|\mathcal{M}_f ^2|}{\mu^2}\right]\,,
    \label{yukawaloop}
\end{equation}
where $n_i=2 (4)$ for a Majorana (Dirac) fermion, and
the coefficient $b_f$ is given by
\begin{equation}
 b_f = -\frac{1}{45}R_{\mu\nu}R^{\mu\nu} -\frac{7}{360}R_{\mu\nu\rho\sigma}R^{\mu\nu\rho\sigma}\simeq -\frac{19}{15}H^4\,.
\end{equation} 
If the curvature effect dominates over the Yukawa coupling, i.e., $H^2\gg y^2\phi^2$, the one-loop potential would be of order $H^4$, which is negligible. Therefore, we are interested in the regime $y^2\phi^2\gg H^2$. If we take $\phi\sim \phi_*\simeq 5.3M_P$, this amounts to $|y|\gg 10^{-6}$ and implys a reheating temperature $\trh \gg 10^8$~GeV. This is the case we consider in the following, so we neglect curvature contributions.

For a first estimate of the one-loop effect, one may consider adding \eqref{yukawaloop} to the Starobinsky potential with a \textit{fixed} coupling constant $y$. The renormalization scale can be chosen such that the $\log$ term is minimized, i.e.,~$\mu=y\phi$. In this case, it is straightforward to derive  $n_s$ in terms of $y$. To satisfy the bound from  \textit{Planck} up to $2\sigma$, we get $y<7.6 \times 10^{-4}$ and from \textit{ACT}, we get $6.4\times 10^{-4}<y<9.9\times 10^{-4}$. However, fixing the  renormalization scale is just a first approximation since the coupling constants are scale-dependent, and obey renormalization group equations (RGEs).

In order to RG-improve the one-loop effective potential \eqref{yukawaloop}, we first compute the one-loop $\beta$-functions of the model parameters, using the Callan-Symanzik (CS) equation:
\begin{equation}
  \left(\frac{\partial}{\partial\log \mu}+\beta_{g_i}\frac{\partial}{\partial g_i}-\gamma _\phi\phi \frac{\partial}{\partial \phi}\right)V_{\text{eff}}(\phi)=0\,, 
  \label{cse}
\end{equation}
with $V_{\text{eff}}(\phi)=V(\phi ) +\Delta V_f(\phi)$. Assuming a Yukawa interaction with a Majorana fermion, the one-loop $\beta$-function of the Yukawa coupling is
\begin{equation}
    \beta_y=\frac{y^3}{4\pi^2}\,.
\end{equation} 
The corresponding RGE has a simple analytic solution:
\begin{equation}
    y(\mu)=\frac{y_0}{\sqrt{1+\frac{1}{2\pi^2}y_0^2\log(\mu_0/\mu )}}\,,
    \label{ysolution}
\end{equation} 
where $\mu_0$ is a reference scale and we denote $y(\mu_0)\equiv y_0$. For $\log(\mu_0/\mu) \sim \mathcal{O}(1)$ and $y_0\sim 10^{-4}$, $y(\mu)$ is almost a constant $y(\mu) \simeq y_0$.
In \eqref{yukawaloop}, the first term gives rise to a $\phi^4$ term in the CS equation, and therefore a higher-order $\lambda \phi^4$ interaction will be generated.~\footnote{We note that if there is a tree-level $\lambda\phi^4$ term in the potential,
it will spoil the flatness of the potential necessary for inflation.} In addition, to cancel the anomalous dimension term, $\gamma_\phi V^\prime \phi $, in the CS equation we also take into account scale dependence in $m^2$ and $M_P$. Solving \eqref{cse}, we obtain the $\beta$-functions
\begin{equation}
\begin{aligned}
    &\beta_{M_P}=-\frac{M_Py^2}{16\pi^2},\quad \beta _{m^2}=\frac{m^2y^2}{8\pi^2 }, \quad
    \beta_\lambda=-\frac{y^4}{16\pi^2} \, ,
    \end{aligned}
\end{equation}
leading to the analytic solutions
\begin{equation}
    \begin{aligned}
&        M_P(\mu)=M_{P0}\left( \frac{y_0}{y(\mu)}\right)^{1/4} ,
\\&m^2(\mu)=m_0^2\left(\frac{y(\mu)}{y_0}\right)^{1/2}
,\\&\lambda(\mu)= \frac{y_0^2y^2(\mu)}{16\pi^2}\log(\mu_0/\mu) \, 
, 
    \end{aligned}\label{solRGE}
\end{equation} 
where the quartic coupling solution assumes $\lambda(\mu_0)=0$. The solutions show
that $M_P$ and $m^2$ depend  only very mildly on the energy scale for the $y$ values of relevance. The inflaton field itself is also scale-dependent, with a renormalization factor
\begin{equation}
    \phi(\mu)=\sqrt{\frac{Z(\mu_0)}{Z(\mu)}}\phi(\mu_0),\quad \gamma_\phi(\mu )\equiv \frac{\partial \log \sqrt{Z(\mu)}}{\partial\log \mu}\,,
    \label{eq:wavefnrenorm}
\end{equation}
where for the Yukawa interaction the anomalous dimension of $\phi$ is given by
\begin{equation}
    \gamma_\phi=\frac{y^2}{16\pi^2}\label{anomphi}\,.
\end{equation}
Thus, using \eqref{anomphi} and \eqref{ysolution} we can also solve \eqref{eq:wavefnrenorm} for $Z(\mu)$ to obtain
\begin{equation}
    Z(\mu)=\left(\frac{y (\mu)}{y_0  }\right)^{1/2}\,,
\end{equation} 
where $Z(\mu_0)=1$. Finally, the RG-improved potential is
\begin{eqnarray}
    V_{RGI}(\phi,\mu)&=&\frac{3}{4}m^2(\mu)M_P^2(\mu)
    \left(1-e^{-\sqrt{\frac{2}{3Z(\mu)}}\frac{\phi}{M_P(\mu)}}\right)^2\nonumber\\
    &&-\,\frac{y^4(\mu)}{32\pi^2}  
    \frac{\phi^4}{Z^2(\mu)}\left(\log\left( \frac{ y^2(\mu)\phi^2 }{\mu^2Z(\mu)}\right)-\frac{3}{2}\right)\nonumber \\
    &&+\,\lambda(\mu) \frac{\phi^4}{Z^2(\mu)}\,.
\label{RGIV}
\end{eqnarray}
We choose as reference scale $\mu_0=M_P$, where the model emerges from UV physics, and $\phi \equiv \phi(\mu_0)$. One can check explicitly that \eqref{RGIV} is independent of $\mu$ up to one-loop order, i.e., taking the derivative with regard to $\mu$, the first line vanishes, and the second plus the third line also vanish up to higher loop terms. This reflects the fact that we have solved the CS equation \eqref{cse} up to one loop, and hence the residual $\mu$-dependence in the RG-improved potential only emerges at two-loop order. 
In principle, the renormalization scale $\mu$ can be chosen so that the second line in \eqref{RGIV} vanishes, which involves evaluating $\mu$ numerically point-by-point.  However, given that the RG-improved potential is independent of $\mu $ up to one loop, we can choose a simpler scale $\mu=y_0 \phi$, close to the mass scale of the fermion.~\footnote{More precisely, to evaluate the 1-loop correction to the potential and the running couplings, it is better to choose $\mu^2 = y_0^2 \phi^2 + V''$. For large $\phi$ (at or near the scale of inflation), this is equivalent to $\mu = y_0 \phi$, but this allows us to determine the corrections near the minimum of the potential where $\mu \simeq V'' \simeq m^2$. Similarly, the inflaton mass is also mostly independent of the renormalization scale.} As we have previously noted, in the chosen parameter range $y_0\sim 10^{-4}$, the running Yukawa coupling is almost constant, i.e., $y(\mu=y_0 \phi) \simeq y_0$.

The RG-improved potential depends on two parameters $y_0$ and $m_{0}$, where
the latter is fixed by the CMB observables. 
Both $y_0$ and $m_{0}$  also appear in the definitions of $N_*$ (see Eqs.~\eqref{e-folds}, \eqref{eq:nstarreh}) and of $A_s$ (Eq.~\eqref{As2}). 
 Each value of $y$ corresponds to a specific reheating temperature, which enters in Eq.~(\ref{eq:nstarreh}), to determine $N_*$.
 Note that the reheating temperature is determined by the low-energy value of $y$ through Eq.~\eqref{eq:TRH}, and $y(\mu=m_0)\simeq y_0$.
These equations can be expressed in terms of    $\phi_*$, $y_0$ and $m_{0}$, with $\phi_{\text{end}}$ replaced by the solution of \eqref{phiendeq}. For any $y_0$, one can then solve numerically the three coupled equations to obtain $m_0$ and $\phi_*$, so that the inflaton potential, as well as all CMB observables are completely determined. In the following we compare these results to the tree-level values using the same procedure to derive $m$ and $\phi_*$ for any given $y_0$.

\begin{figure}[h]
    \centering
    \includegraphics[width=.48\textwidth]{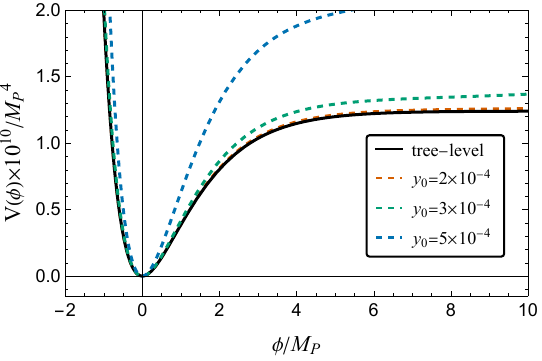}
    \caption{The RG-improved effective potential as a function of $\phi$ for the indicated values of $y_0$, compared to the tree-level potential.  }
    \label{efffermion}
\end{figure}

In Fig.~\ref{efffermion} we display the one-loop-corrected effective potential due to the coupling of a Majorana fermion to the Starobinsky inflaton. The dashed lines correspond to the RG-improved potential given in \eqref{RGIV} with the renormalization scale  $\mu=y_0 \phi$ for three representative values of $y_0$. For the tree-level potential (black solid line), we have taken $m\simeq 1.3\times10^{-5}M_P $ corresponding to $y_0 = 9.5\times 10^{-5}$, and  $\trh\simeq 4  \times 10^{10} \text{ GeV}$.

\begin{figure}[h]
    \centering
    \includegraphics[width=.48\textwidth]{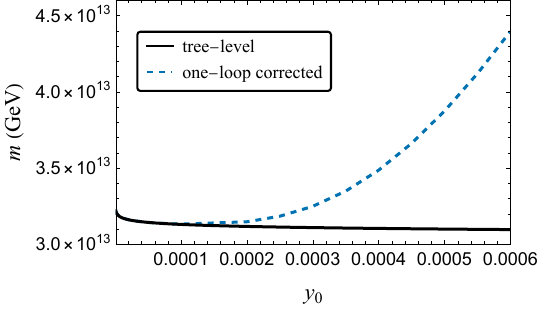}
    \caption{The inflaton mass $m$ as a function of $y_0$, as obtained from the tree-level potential \eqref{treestaro} (black solid line) and from the one-loop corrected potential \eqref{RGIV} (blue dashed line). }
    \label{minfl}
\end{figure}

\begin{figure}[h]
    \centering
    \includegraphics[width=.48\textwidth]{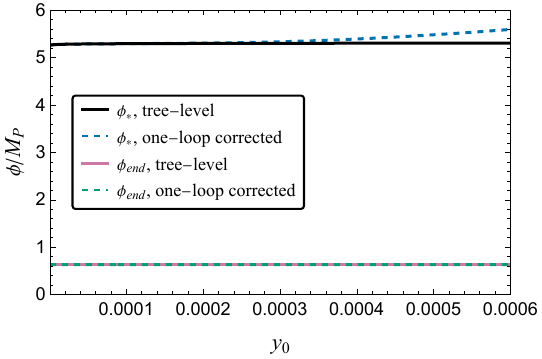}
    \caption{The inflaton field values  $\phi_*$ and $\phi_{\text{end}}$ as functions of $y_0$, from the tree-level potential \eqref{treestaro} (solid lines) and from the one-loop corrected potential \eqref{RGIV} (dashed lines). }
    \label{phicomp}
\end{figure}

In order to study the effects of one-loop corrections on the inflaton potential, we first show in Fig.~\ref{minfl} the inflaton mass as a function of $y_0$, where the inflaton mass is obtained from the  one-loop corrected potential by evaluating
$m ^2=\partial^2_{\phi\phi}V_{RGI}(\phi,y_0\phi)$ at $\phi = 0$. For $y_0\lesssim10^{-4}$, the one-loop correction is negligible, whereas for larger $y_0$, the correction of $m$ becomes increasingly significant. This effect can be seen also in Fig.~\ref{efffermion}, where the potential around the origin is steeper for larger $y_0$. We show in Fig.~\ref{phicomp} the corresponding comparison between the tree-level and one-loop results for $\phi_*$ and $\phi_{\text{end}}$, where we see that the one-loop corrections are much smaller.

\begin{figure}[h]
    \centering
    \includegraphics[width=.48\textwidth]{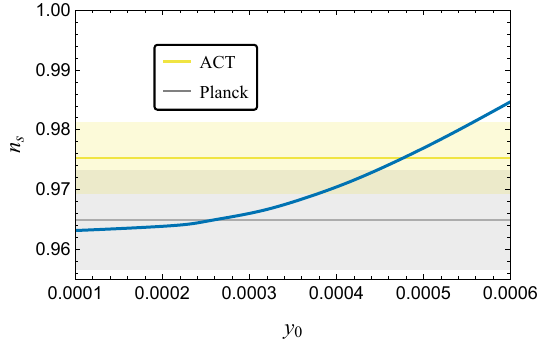}
    \caption{The spectral index $n_s$ from the one-loop corrected potential as a function of the Yukawa coupling $y_0$. The gray shaded region is the constraint on $n_s$ from {\it Planck} with $\pm 2\sigma$ uncertainty, while the yellow band is the ACT constraint.}
    \label{nsfermion}
\end{figure}

The resulting scalar spectral index for $y_0$ in the range $[10^{-4},6\times 10^{-4}]$ is shown in Fig.~\ref{nsfermion}. Consistency with the {\it Planck} result  within $2\sigma$ imposes an upper bound for the Yukawa coupling: $y< 4.5\times 10^{-4}$, which corresponds to an upper bound for the reheating temperature $T_{\text{RH}}< 2  \times 10^{11} \text{ GeV}$. If we consider instead the ACT constraint on $n_s$, the Yukawa coupling is constrained to be $3.8\times10^{-4}< y< 5.6\times10^{-4}$, corresponding to the reheating temperature constraint $1.7 \times 10^{11}<\trh< 2.8 \times 10^{11} \text{ GeV}$. 

\begin{figure}[h]
    \centering
    \includegraphics[width=.48\textwidth]{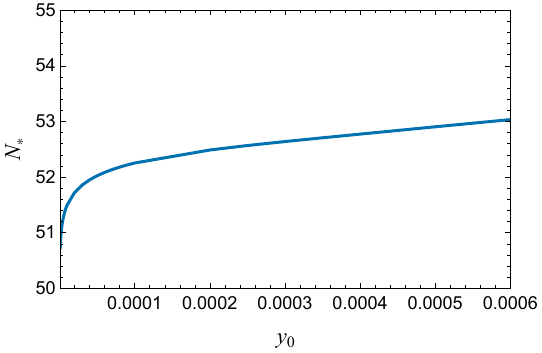}
    \caption{The number of $e$-folds $N_*$ from the one-loop corrected potential as a function of the Yukawa coupling $y_0$.}
    \label{nstarfermion}
\end{figure}

\begin{figure}[htbp]
    \centering
    \includegraphics[width=.48\textwidth]{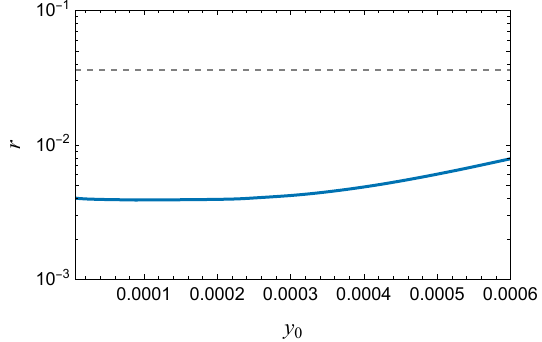}
    \caption{The tensor-to-scalar ratio from the one-loop corrected potential as a function of the Yukawa coupling $y_0$.  The gray dashed line is the upper bound \eqref{rlim} from {\it Planck}/BICEP/Keck bound \cite{BICEP2021}.
    }
    \label{rfermion}
\end{figure}

We show in Fig.~\ref{nstarfermion} the number of $e$-folds in terms of the Yukawa coupling. At the upper bound $y=4.5\times 10^{-4}$, we have $N_*\simeq53$.
In Fig.~\ref{rfermion} we show the tensor-to-scalar ratio from the one-loop corrected potential. The {\it Planck}/BICEP/Keck bound \cite{BICEP2021} imposes $r<0.036$ which is always satisfied in the considered range of $y$.

\subsubsection{Massive fermion $m_f\neq 0$}

The fermion could in principle have a non-negligible mass term in the Lagrangian, in which case we include the bare mass $m_f$ (though still neglecting $H$). The running mass $m_f(\mu)$ is determined by \cite{Luo:2002ti}:
\begin{equation}
    \beta_{m_f}=\frac{y^2m_f}{4\pi^2}   \, .
\end{equation}
Using \eqref{ysolution}, we obtain an analytic solution:
\begin{equation}
    m_f(\mu)=m_{f_0}\frac{y(\mu)}{y_0} \,.
\end{equation}
With a nonzero mass term, one must add the contributions $\Lambda+a_1\phi +a_2\phi^2+a_3\phi^3 +\lambda\phi^4$ to the potential for the RGEs to close, where $\Lambda$ is a constant energy density. Similar to the massless case, solving the CS equation \eqref{cse} yields analytic solutions for the running parameters, namely
\begin{equation}
    \begin{aligned}
        &\Lambda(\mu)=\frac{m_{f_0}^2m_f^2(\mu)}{16\pi^2}    \log(\mu_0/\mu)+\Lambda_0  \, , \\& a_1(\mu)=\frac{y(\mu)m_{f_0}^2m_f(\mu)}{4\pi^2}\log(\mu_0/\mu)\, , 
        \\&a_2(\mu)=\frac{3 m_f^2(\mu)y_0^2}{8\pi^2}\log(\mu_0/\mu)  \, , \\&a_3(\mu)=\frac{y_0^2y(\mu) m_f(\mu)}{4\pi^2}\log(\mu_0/\mu)  \, , 
    \end{aligned}
\end{equation}
where the parameters $  a_1,a_2,a_3$ are required to vanish  at the scale $\mu_0$ and $\Lambda_0$ is the vacuum energy at the UV scale $\mu_0$. The other running parameters $M_P(\mu)$, $m^2(\mu)$, $\lambda(\mu)$ are still given by \eqref{solRGE}.  

\begin{figure}[h]
    \centering
    \includegraphics[width=.48\textwidth]{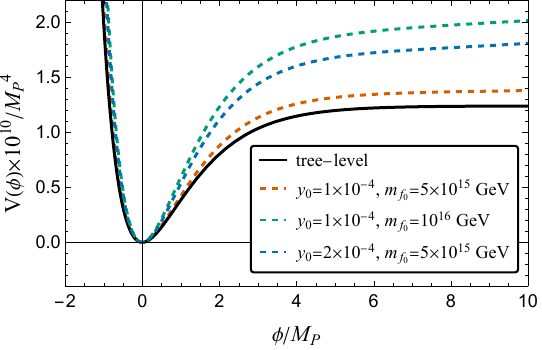}
    \caption{The RG-improved effective potential as a function of $\phi$ for various values of $y_0$ and $m_{f_0}$, compared to the tree-level potential.
    }
    \label{efffermionmf}
\end{figure}

The RG-improved potential for a massive fermion then becomes
\begin{eqnarray}
\label{RGImf}
    V_{RGI}(\phi,\mu)&=&\frac{3}{4}m^2(\mu)M_P^2(\mu)
    \left(1-e^{-\sqrt{\frac{2}{3Z(\mu)}}\frac{\phi}{M_P(\mu)}}\right)^2 \nonumber\\
    &&-\,\frac{\mathcal{M}_f^4(\mu)}{32\pi^2}  \left(\log\left( \frac{ \mathcal{M}_f^2(\mu) }{\mu^2}\right)-\frac{3}{2}\right)\nonumber \\
    &&+\,\lambda(\mu) \frac{\phi^4}{Z^2(\mu)}+a_1(\mu)\frac{\phi}{\sqrt{Z(\mu)}}+a_2(\mu)\frac{\phi^2}{Z(\mu)}\nonumber\\
    &&+\,a_3(\mu)\frac{\phi^3}{Z(\mu)^{3/2}}+\Lambda(\mu)\,,
\end{eqnarray}
where 
\begin{equation}
    \mathcal{M}_f(\mu)=y(\mu) \frac{\phi}{\sqrt{Z(\mu)}}+m_f(\mu)\,.
\end{equation}
To analyze the effect of $y_0$ and $m_{f_0}$ on the CMB predictions, we will fix the initial value $\Lambda_0$, such that $V_{RGI}(0,\mu)=0$. 
In the computation of the one-loop correction, we have assumed that the Hubble constant $H$, as well as its time variation $\dot{H}$ are negligible. This is a good approximation for the chosen parameter space. Since $m_f(\mu)$ and $y(\mu)$  vary slowly with $\mu$, we can choose the renormalization scale    $\mu=y_0\phi+m_{f_0}$.   In Fig.~\ref{efffermionmf}, we show the one-loop corrected potential compared to the same tree-level potential shown in Fig.~\ref{efffermion}, for different values of $y_0$ and $m_{f_0}$.

\begin{figure}[h]
    \centering
    \includegraphics[width=.48\textwidth]{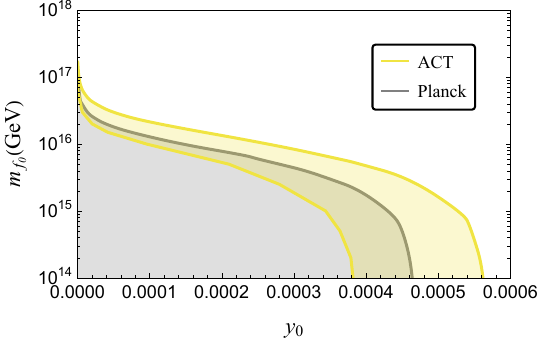}
    \caption{The $(y_0, m_{f_0})$ contours  satisfying the $n_s$ constraints from   \textit{Planck}   (gray shaded region), and {ACT} (yellow shaded region), up to $2\sigma$.  } 
    \label{mfcontour}
\end{figure}

In Fig.~\ref{mfcontour}, we show the $(y_0,m_{f_0})$ parameter space that is allowed by the  \textit{Planck} and \textit{ACT} constraints on $n_s$. For $y_0= 10^{-4}$, \textit{Planck} implies an upper bound $m_{f_0} <1.2\times  10^{16} \text{ GeV}$, whereas \textit{ACT} constrains the bare mass to be in the range $ 10^{16}\text{ GeV} <m_{f_0}<2\times 10^{16} \text{ GeV}$. When $y_0\rightarrow0 $, the fermion decouples from the inflaton and therefore does not contribute to the one-loop correction. In this case, there is no constraint on $m_{f_0}$. Note that when the fermion mass is larger than the inflaton mass, the decay channel $\phi\rightarrow f f$ is kinematically forbidden, implying the Yukawa coupling $y$ is no longer related to the reheating temperature.
For the numerical evaluations, we have considered a reheating temperature  $\trh= 4\times 10^{10}\text{ GeV}$.

\subsection{Yukawa interaction in the Jordan frame}\label{fermjordan}

As an alternative to the fermion effective mass generated by the Yukawa interaction in the Einstein frame discussed in Sec.~\ref{sec:FermionEinstein}, we now consider a possible coupling to the curvature in the Jordan frame, analogous to the non-minimal scalar coupling $\xi R \varphi^2$. We adopt the following action in the Jordan frame:
\begin{equation}
    \begin{aligned}
S= \int d^4 x \sqrt{-\widetilde{g}}&\left( -\frac{M_P^2}{2}\widetilde{R}+\alpha \widetilde{R}^2 +\frac{i}{2}\bar{f}\slashed{D} f\right. \\&\quad   \left.-\frac{1}{2}m_f\bar{f}f+\beta \widetilde{R} \bar{f}f+\frac{1}{4} \frac{\beta^2}{\alpha} \bar{f}f\bar{f}f\right)\,,
    \end{aligned}
\label{jordferm}
\end{equation}
where $\beta $ is a non-minimal coupling with mass dimension $-1$, and the four-fermion coupling is chosen to have the value $\beta^2/\alpha$. 
The covariant derivative acts on the fermion as $D_\mu f=\partial_\mu f -\frac{1}{4}\omega _\mu ^{ab}[\gamma_a,\gamma_b]f$ where $\omega _\mu ^{ab}$ is the spin connection.  This action is equivalent to:
\begin{equation}
    \begin{aligned}
        S=\int d^4 x\sqrt{-\widetilde{g}} &\left(-\frac{M_P^2}{2}\widetilde{R} +2\alpha \Phi \widetilde{R}-\alpha \Phi^2\right. \\&\quad\left.-\frac{1}{2}m_f\bar{f}f+\beta \Phi \bar{f}f+\frac{i}{2}\bar{f}\slashed{D} f\right) \, .
    \end{aligned}
\label{jordint}
\end{equation}
The equation of motion for the auxiliary scalar  is $\Phi=\widetilde{R}+\frac{\beta}{2\alpha} \bar{f}f$. Once  $\Phi$ is integrated out, we recover \eqref{jordferm}. 

Starting from \eqref{jordint}, we perform the conformal transformation  (\ref{gconf}) where again $\Omega$ is determined by (\ref{Oconf}). We note that the kinetic term of the fermion is invariant under this transformation:
\begin{equation}
    \int d^4 x\sqrt{-\Tilde{g}}\frac{i}{2}\bar{f}\slashed{D} f=\int d^4 x\sqrt{-g}\frac{i}{2}\bar{F}\slashed{\tilde{D}} F \, ,
\label{confferm}
\end{equation}
with the relation $F=fe^{-\frac{3}{2}\Omega}$. Eq.~\eqref{confferm} is the well-known result that a massless spin $1/2$ fermion is conformally invariant, see, e.g.,~\cite{Parker:2009uva}. However, the conformal invariance is explicitly broken when $m_f,\beta\neq0 $. After the redefinition $\phi=\sqrt{6}M_P \Omega$, we obtain the following action in the Einstein frame:~\footnote{From the Einstein-frame Lagrangian, we see that the fermion scattering amplitude is well-behaved at high energy provided $\beta \lesssim M_P/m^2$ . This behavior, which is not manifest in the Jordan frame Lagrangian, follows from the specific four-fermion coupling in \eqref{jordferm}.}
\begin{equation}
    \begin{aligned}
S= \int& d^4 x  \sqrt{-g}\left[  -\frac{M_P^2}{2}\Tilde{R}+\frac{i}{2}\bar{F}\slashed{\Tilde{D}} F+\frac{1}{2}\Tilde{\partial}_\mu\phi\Tilde{\partial}^\mu\phi-V(\phi)\right. \\&   \left.-\frac{1}{2}m_f e^{-\frac{\phi}{\sqrt{6}M_P}}\bar{F}F-\frac{M_P^2\beta}{2\alpha }\sinh \left(\frac{\phi}{\sqrt{6}M_P}\right)\bar{F}F\right] \, ,
    \end{aligned}
\end{equation}
where $V(\phi)$ is the Starobinsky potential given in (\ref{treestaro}). For small field values $\phi\ll M_P$, the last term is approximately a Yukawa coupling $\propto \phi \bar{F}F$. On the other hand, the fermion mass term $m_f \bar{f}f$ in the Jordan frame action \eqref{jordferm}, results in a coupling that is exponentially suppressed in the Einstein frame  at large inflaton field value. However,   the one-loop correction from this coupling may still be significant for large $m_f$.   In this case, we can derive an upper bound on the fermion mass.

In order to evaluate the radiative corrections, we perform the same analysis as before, using the field-dependent effective fermion mass:
\begin{equation}
   \mathcal{M}_f= m_fe^{-\frac{\phi}{\sqrt{6}M_P}}+\frac{\beta}{\alpha}M_P^2\sinh \left(\frac{\phi}{\sqrt{6}M_P}\right) \, .
\end{equation}
We assume the couplings dominate over curvature effects, allowing us to neglect the $H^2$ piece in the effective mass.
The effective Yukawa coupling is
\begin{equation}
    y(\phi)=- \frac{m_f}{\sqrt{6}M_P} e^{-\frac{\phi}{\sqrt{6}M_P}}+\frac{M_P \beta}{ \sqrt{6}\alpha }\cosh \left(\frac{\phi}{\sqrt{6}M_P}\right) \,.
\end{equation}
 In principle, we can follow the same RG-improvement procedure as in Section~\ref{1loopf}, but the counterterms are more complicated due to the non-renormalizable couplings. For simplicity, we estimate the one-loop correction using fixed couplings, with $\beta=0$, and we choose $\mu=\mathcal{M}_f (\phi)$.  This choice makes the one-loop correction \eqref{yukawaloop} small around $\phi_*$, where we evaluate the CMB observables. Since the potential is not RG-improved, the range of validity of the potential is limited, and cannot be extended to $\phi\ll\phi_*$.   For an estimate of the limit, we fix $m=1.3 \times 10^{-5}M_P$ for the tree-level potential, and use \eqref{As2} to evaluate $\phi_*$. 
This leads to the bounds 
$m_f\lesssim5.8\times 10^{16}\text{ GeV}$ from \textit{Planck} and $5.4 \times 10^{16}\lesssim m_f\lesssim6.2\times 10^{16}\text{ GeV}$ from \textit{ACT}. The assumption of fixed couplings is a good approximation, since in the $m_{f_0}=0$ case, the bounds obtained with fixed and running couplings are nearly identical.

\section{One-Loop Corrections from Inflaton Couplings to Scalars}
\label{1loopb}
\subsection{Trilinear coupling to the inflaton}

In this section we consider a real scalar field $s$  coupled to the inflaton $\phi$, including the possibility that the scalar has a non-minimal coupling $\xi$ to gravity, in which case the scalar part of the action is
\begin{equation}
    S_s=\int d^4 x \sqrt{-g} \left[\frac{1}{2}g^{\mu\nu}\partial_\mu s\partial_\nu s-\frac{1}{2}\xi R s^2  -V(s,\phi)\right] \, .
\end{equation}
Perturbing the scalar around its background value $s_0$, i.e., $s=s_0+\hat{s}$, one finds that the quadratic terms for the perturbation $\hat{s}$ are
\begin{equation}
    S_s=\frac{1}{2}\int d^4 x \sqrt{-g} \left[ -\hat{s}\left(\Box+\xi R+\partial_{ss}^2V(s,\phi)\right)\hat{s}+\cdots\right] \, .
\end{equation}
The fluctuation operator is $\Box +\partial_{ss}^2V(s,\phi)$, and the effective scalar mass is
\begin{equation}
    \mathcal{M}_s^2= \partial_{ss}^2V(s,\phi)\Big|_{s_0} -\left(\xi-\frac{1}{6}\right)R \, ,
\end{equation}
and the scalar contribution to the one-loop effective potential\footnote{{A quadratically divergent one-loop tadpole term has been tuned to zero and, as in the fermion case, we focus solely on the logarithmically divergent terms.}} is:
\begin{equation}
V^{(1)}_s(\phi)=    \frac{n_i}{64\pi^2} \left[ \mathcal{M}_s ^4 \left(\log \frac{ |\mathcal{M}_s ^2| }{\mu^2}-\frac{3}{2}\right)+2 b_s \log \frac{|\mathcal{M}_s^2| }{\mu^2}\right]\,,
\label{loopscalar2}
\end{equation}
with
\begin{equation}
       b_s=-\frac{1}{180}R_{\mu\nu}R^{\mu\nu} +\frac{1}{180}R_{\mu\nu\rho\sigma}R^{\mu\nu\rho\sigma} \, , \label{bs}
\end{equation}
where $n_i$ is the number of scalar degrees of freedom ($n_i=1$ for one real scalar). We consider a real scalar possessing a bare mass $m_s$ and interacting with the inflaton through the trilinear term $\frac{1}{2}\kappa \phi s^2$, where $\kappa$ is a dimensionful coupling.
The effective mass is then
\begin{equation}
    \mathcal{M}_{s}^2=\kappa \phi+m_s^2-\left(\xi -\frac{1}{6}\right)R\,.
\end{equation}
In the de Sitter case ($R=-12 H^2$) and for  $\xi\sim \mathcal{O}(1)$, we can neglect the curvature for $\kappa\gg 10^7 \text{ GeV}$ or $m_s\gg 10^{13}\text{ GeV}$. This approximation may change for large $\xi$. However, since its contribution to the effective potential scales as $\xi^2 H^4$, which is roughly $10^{-10}\xi^2 \times V(\phi)$, this term will affect inflationary observables only for very large $\xi \sim 10^4-10^5$. In the following, we work in the large $\kappa$ (or $m_s$) and small $\xi$ regime, so that the flat spacetime approximation applies.

\subsubsection{Massless scalar $m_s=0$}
As in the fermion case, we first consider a scalar without a bare mass.~\footnote{By this we mean a bare mass negligible compared to the inflation scale but not so small as to trigger a vacuum expectation value (VEV) for the scalar.} 
Assuming that reheating is dominated by the decay channel $\phi\rightarrow ss$, we obtain the reheating temperature as follows:
\beq
g_{\rm RH} \frac{\pi^2}{30} \trh^4 = \frac{12}{25} (\Gamma_\phi M_P)^2 =\frac{3M_P^2\kappa^4}{6400\pi^2 m^2 }  \,.
\eeq
At one loop, no anomalous dimension is generated by the cubic $\phi s^2$ coupling, and $\beta_\kappa=0$ since the one-loop diagram with $\phi ss$ external legs is convergent. Therefore, the CS equation is
\begin{equation}
  \left(\frac{\partial}{\partial\log \mu}+\beta_{g_i}\frac{\partial}{\partial g_i} \right)V_{\text{eff}}(\phi)=0\,, 
  \label{cse1}
\end{equation}
with $V_{\text{eff}}=V+V^{(1)}_s$. The first term in \eqref{cse1} gives rise to a $\kappa^2 \phi^2 $ term, which will be canceled by a counterterm $b_2\phi^2$ that appears only radiatively.
We then obtain the $\beta$-functions
\begin{equation}
    \beta_{b_2}=\frac{\kappa^2}{32\pi^2},\quad \beta_{M_P}=\beta _{m^2}=\beta_\kappa=0\,,
\end{equation}
which yield 
\begin{equation}
    b_2(\mu)=\frac{\kappa^2}{32\pi^2}  \log(\mu/\mu_0)\,. 
\end{equation}
As in the fermion case, we choose the reference scale $\mu_0=M_P$ and the renormalization scale is taken to be the scalar effective mass $\mu=\sqrt{\kappa \phi }$.\footnote{As in the fermion case, for low $\phi$ (i.e. $\kappa \phi < m^2$) we must include the inflaton mass and take $\mu^2 = \kappa \phi + V''$.} 
The RG-improved potential then becomes
\begin{eqnarray}
        V_{RGI}(\phi, \mu)&=&\frac{3}{4}m^2M_P^2\left(
            1-e^{-\sqrt{\frac{2}{3}}\frac{\phi}{M_P}}
        \right)^2+b_2(\mu)\phi^2\nonumber\\
        &&+\,\frac{1}{64\pi^2}\kappa^2\phi^2\left(\log \frac{\kappa\phi}{\mu^2}-\frac{3}{2}\right)\,.
\label{VRGIscalar}
\end{eqnarray}

\begin{figure}[h]
    \centering
    \includegraphics[width=.48\textwidth]{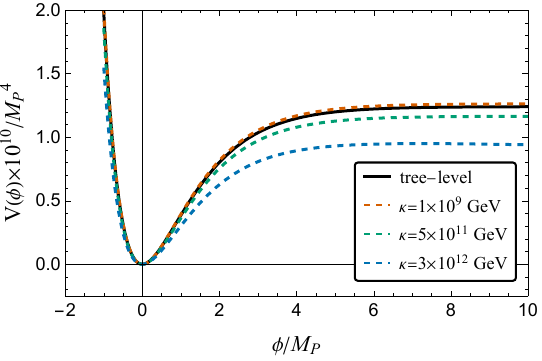}
    \caption{The RG-improved effective potential as a function of $\phi$ for the indicated values of the trilinear scalar coupling $\kappa$, compared to the tree-level potential. }
    \label{effscalar}
\end{figure}

\begin{figure}[h]
    \centering
    \includegraphics[width=.45\textwidth]{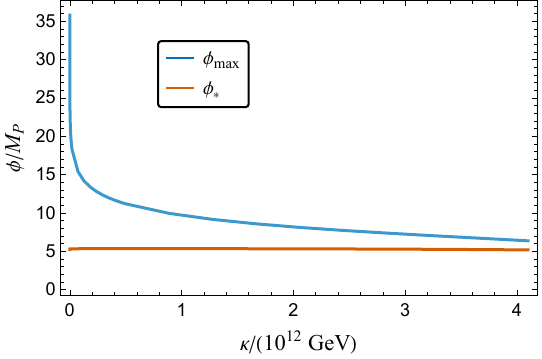}
    \caption{The values of $\phi_{max}$, where the one-loop potential has a local maximum, and $\phi_*$ as functions of the trilinear scalar coupling $\kappa$.}
    \label{phimaxscalar}
\end{figure}

In Fig.~\ref{effscalar} we compare the RG-improved potential \eqref{VRGIscalar} to the tree-level potential (with $m=1.3\times   10^{-5}M_P,\trh=4\times 10^{10}\text{ GeV}$), for different values of $\kappa$.
We note that the one-loop radiative correction from the scalar causes the slope of the inflaton potential to turn negative at large field values. For example, if $\kappa=4\times 10^{12}\text{ GeV}$ this happens at $\phi_{max}=6.4 M_P$, to be compared with the value of the inflaton field at the pivot scale, $\phi_*=5.2 M_P$. In this case, inflation cannot start from an initial value $\phi\geq 6.4 M_P$, if the inflaton is initially at rest. In Fig.~\ref{phimaxscalar} we show the  field value $\phi_{max}$, where the potential has a local maximum, and $\phi_*$ as functions of $\kappa$.

\begin{figure}[h]
    \centering
    \includegraphics[width=.45\textwidth]{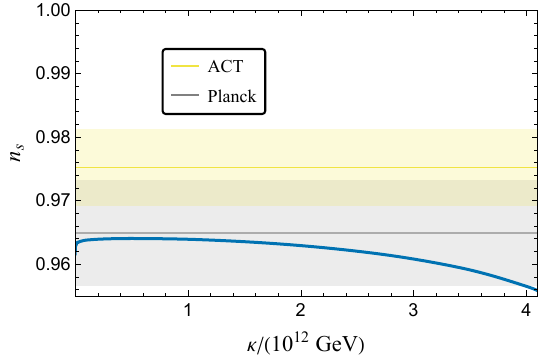}
    \caption{The spectral index $n_s$ obtained from the one-loop corrected potential as a function of the trilinear scalar coupling $\kappa$. The gray shaded region is the constraint on $n_s$ from {\it Planck} with $\pm 2\sigma$ uncertainty, and the yellow band is the ACT constraint.}
    \label{nsscalar}
\end{figure}

\begin{figure}[h]
    \centering
    \includegraphics[width=.45\textwidth]{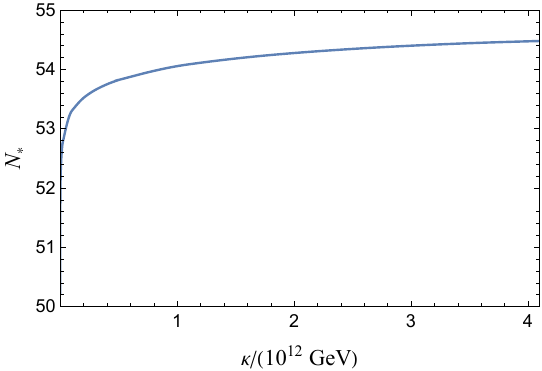}
    \caption{The number of $e$-folds $N_*$ obtained from the one-loop corrected potential as a function of the trilinear scalar coupling $\kappa$. }
    \label{nstarscalar}
\end{figure}

We show in Fig.~\ref{nsscalar} the spectral index $n_s$  as a function of $\kappa$.  \textit{Planck} 2018 data set an upper limit  $\kappa \leq 4\times 10^{12}\text{ GeV}$, implying $\trh\leq 4.2 \times 10^{13}\text{ GeV}$, but $n_s$ never reaches the ACT range.  Within the chosen parameter range, the tensor-to-scalar ratio $r\sim 4\times10^{-3} $,   satisfying the \textit{Planck} upper bound. The number of $e$-folds as a function of $\kappa$ is shown in Fig.~\ref{nstarscalar}.


\subsubsection{Massive scalar: $m_s \neq 0$}
\begin{figure}[h]
    \centering
    \includegraphics[width=.48\textwidth]{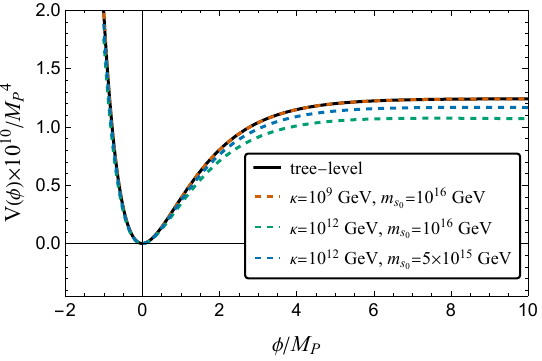}
    \caption{The RG-improved effective potential as a function of $\phi$ for the indicated values of the trilinear scalar coupling $\kappa$ and bare mass $m_{s_0}$, compared to the tree-level potential. }
    \label{effscalarbare}
\end{figure}

\begin{figure}[h]
    \centering
    \includegraphics[width=.46\textwidth]{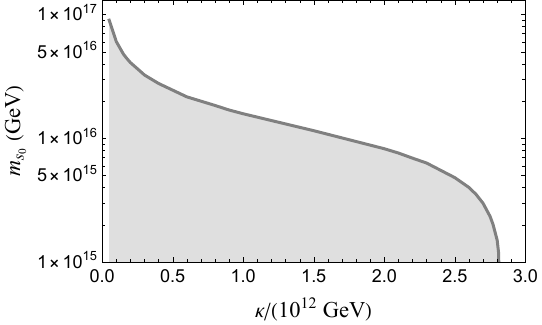}
    \caption{The $(\kappa, m_{s_0})$ contour within which the $n_s$ constraint from \textit{Planck} is satisfied at the $2\sigma$ level. The bound from \textit{ACT} DR6 cannot be satisfied for any value of $\kappa$.} 
    \label{mscontour}
\end{figure}

We consider next a scalar with non-vanishing bare mass $m_s\neq 0$. At one-loop order, the bare mass term receives a contribution from the trilinear coupling, with $\beta$-function
\begin{equation}
    \beta_{m_s^2}=\frac{\kappa^2}{8\pi^2} \, ,
\end{equation}
which leads to the exact solution:\begin{equation}
m_s^2(\mu)=\frac{\kappa^2}{8\pi^2}\log(\mu/\mu_0)+m_{s_0}^2 \, ,
\end{equation}
where $m_{s_0}\equiv m_s^2(\mu_0)$. 
In our implementation of Eq.~\eqref{cse1}, we introduce the counterterms $
b_1(\mu)\phi+b_2(\mu)\phi^2 +\Lambda(\mu)$ with:
\begin{equation}
\begin{aligned}
& \Lambda (\mu)= \frac{1}{32\pi^2}m_{s_0}^4 \log(\mu/\mu_0)+\Lambda_0 \, ,\\
& b_1(\mu)=\frac{1}{16\pi^2}   m_{s_0}^2\kappa\log(\mu/\mu_0)   \, , 
\\
& b_{2}(\mu)=\frac{1}{32\pi^2}\kappa^2\log(\mu/\mu_0) \,.  
 \end{aligned}
\end{equation}
We choose the renormalization scale,  $\mu=\sqrt{m_{s_0}^2+\kappa \phi}$, and find the following RG-improved potential:
\begin{eqnarray}
\label{RGImscalar}
    && \hspace{-5mm} V_{RGI}(\phi,\mu) \nonumber\\
    &=&\frac{3}{4}m^2 M_P^2 
    \left(1-e^{-\sqrt{\frac{2}{3 }}\frac{\phi}{M_P }}\right)^2 +b_1(\mu )\phi +b_{2}(\mu)\phi^2\nonumber\\
    &&+\,\frac{1}{64\pi^2}(\kappa\phi+m_{s_0}^2)^2\left(\log \frac{\kappa\phi+m_{s_0}^2}{\mu^2}-\frac{3}{2}\right)+ \Lambda(\mu) \nonumber\, .\\
\end{eqnarray}
As in the fermionic case, we impose $V_{RGI}(0,\mu)=0$ to fix $\Lambda_0$. Since we consider heavy $m_s$, the decay channel $\phi\rightarrow ss$ is kinematically forbidden, and therefore it cannot be responsible for reheating. For definiteness, as in previous examples, we fix the reheating temperature to   $\trh=4\times 10^{10}\text{ GeV}$ in the following. The RG-improved potential for different values of $\kappa$ and $m_{s_0}$ is shown in Fig.~\ref{effscalarbare}.~\footnote{In the allowed parameter space, and for $\phi=0$, we find $m_s \simeq m_{s_0}$.} 
In Fig.~\ref{mscontour}, we show the  $(\kappa, m_{s_0})$ contour satisfying the bound on $n_s$ from \textit{Planck}. The bound from \textit{ACT} DR6 cannot be satisfied for any choice of the parameters. 

\subsection{Scalar coupling in the Jordan frame}

Finally, we consider a real scalar field whose Lagrangian is given by 
\begin{align}
    {\cal L}_{\rm matter} &\supset {\cal L}_{\rm scalar}
    = \frac{1}{2}\widetilde g^{\mu\nu}\partial_\mu\chi\partial_\nu\chi - V(\chi) \, .
\end{align}
Redefining $\chi = e^\Omega s$, we may write the kinetic term in canonical form:
\begin{align}
    e^{-4\Omega}{\cal L}_{\rm scalar} &=
    \frac{1}{2}\partial^\mu s\partial_\mu s + \frac{1}{\sqrt{6}M_P}s\partial_\mu s\partial^\mu\phi \nonumber\\
    & + \frac{1}{12}\frac{1}{M_P^2}s^2\partial_\mu\phi\partial^\mu\phi - e^{-4\Omega} V(e^{\Omega}s) \, .
    \label{eq:normalized matter L}
\end{align}
We note that the derivative coupling of $\phi$ does not affect the inflaton potential, as it has shift symmetry.
On the other hand, the potential terms may or may not affect the flatness of the inflaton potential through loop corrections.
For instance, if we simply consider a non-interacting scalar in the Jordan frame with
\begin{align}
    V(\chi) &= \frac{1}{2}m_\chi^2\chi^2 
    \, ,
\end{align}
we find
\begin{align}
    e^{-4\Omega} V(e^{\Omega}s) &=
    \frac{1}{2}(e^{-\Omega} m_\chi)^2 s^2
    \, .
\end{align}
Since $\Omega$ increases linearly with $\phi$, the couplings with nonzero mass dimension  also increase.

One may eliminate the cubic coupling with derivatives in Eq.~(\ref{eq:normalized matter L}) by making a further field redefinition.
Replacing $\phi$ and $s$ as follows:
\begin{align}
    \phi &\to \phi - \frac{1}{2\sqrt{6}M_P}s^2 - \frac{1}{12M_P^2}s^2\phi \, ,\\
    s &\to s + \frac{1}{36M_P^2}s^3 - \frac{1}{12M_P^2}s\phi^2 \, ,
    \label{eq:transformation}
\end{align}
we obtain
\begin{align}
    {\cal L}_\phi + {\cal L}_s &\supset
    -\frac{1}{12M_P^2}s^2(\partial_\mu s)^2 - \frac{1}{3M_P^2}s\phi \partial_\mu s\partial^\mu\phi \, .
    \label{eq:quarticmixing}
\end{align}
We note that the kinetic terms for $\phi$ and $s$ remain canonical at leading order after this transformation.
The effective mass of $s$ is:
\begin{align}
    \mathcal{M}_s^2(\phi)&= (e^{-\Omega} m_\chi)^2(1-\frac{1}{2} \Omega^2)^2\nonumber\\
     &- \frac{1}{2}(e^{-\Omega} m)^2(1-e^{-2\Omega})(1+\Omega),
\end{align} 
To estimate the constraint on $m_\chi$, we follow the same approach as in Sec.~\ref{fermjordan}, with fixed coupling constants. 
For the tree-level potential, we take $m=1.3 \times 10^{-5}M_P$. 
We assume that $\chi$ does not develop a VEV, and therefore we require $\mathcal{M}_s^2\geq 0$.
For $\phi \sim 5 (10)M_P$ this implies $m_\chi\gtrsim 1.5 \times 10^{-5}(2.8\times 10^{-6})M_P$.\footnote{We note that lighter fields may develop a VEV, but these fields are all subject to large scale field fluctuations, see e.g.,~\cite{Garcia:2025rut}. After inflation, at small $\phi$, we have $\mathcal{M}_s^2 =  m_\chi^2$.} In order to satisfy the \textit{Planck} bound on $n_s$, we require $m_\chi\lesssim 0.024 M_P$, whereas the \textit{ACT}  range will never be obtained with scalar radiative corrections. Interestingly, this limit is effectively an upper limit on GUT-scale physics if introduced in Jordan frame, and does not depend on any additional coupling, such as $\kappa$, between the inflaton and scalar field.

\section{Effects of Supersymmetry}
\label{susy}

The effective potential calculations discussed above arise from vacuum insertion diagrams. Combining the contributions of fermions and bosons, one has
\begin{equation}
\Delta V=    \frac{1}{64\pi^2} Str\Bigg\{ \mathcal{M}_i ^4 \left(\log \frac{ |\mathcal{M}_i ^2| }{\mu^2}-\frac{3}{2}\right)+2 b_i \log \frac{|\mathcal{M}_i^2| }{\mu^2} \Bigg\} \, ,
\label{loopscalar-fermion}
\end{equation}
where $Str\{...\}$ denotes a sum over all degrees of freedom, with factors of $+1 (-1)$ for each bosonic (fermionic) degree of freedom.

It has been argued that inflation cries out for supersymmetry~\cite{Ellis:1982ed}, and that the most appropriate framework is no-scale supergravity~\cite{Ellis:1984bf}: for an example of a no-scale supergravity realization of the Starobinsky model, see~\cite{Ellis:2013xoa}. In a supersymmetric model, the number of bosonic and fermionic degrees of freedom are equal, as are their masses. Therefore there would be no analogue of the constraints discussed above on the possible couplings of the inflaton to fermions and bosons. However, supersymmetry is broken at some scale $\tilde m$, which should be taken into account, leading to contribution to the effective potential of the form
\begin{equation}
\Delta V=    {\cal O}\left(\frac{1}{64\pi^2} \tilde m^2 \phi^2 \right) \, .
\label{loopscalar-fermion-broken}
\end{equation}
A specific U(1) supersymmetry-breaking model containing two Higgs supermultiplets with charges $\pm 2$, in which the inflaton was identified with a linear combination of their scalar components (that does not realize Starobinsky-like inflation), was studied in~\cite{Nakayama:2011ri}, with the result
\begin{equation}
    \Delta V \simeq \frac{g^2}{8\pi^2} \left( 1 - 3 \ln \frac{g^2 \phi^2}{\mu^2} \right) \tilde m^2 \phi^2 \, ,
    \label{NT}
\end{equation}
at one-loop order and neglecting higher orders in $\tilde m/g \phi$. In the case of the Einstein-frame inflaton-fermion coupling discussed above, the maximum contribution to $V$ allowed by {\it Planck} data when $\phi \sim 5 M_P$ is ${\cal O}(10^{-11})M_P^4$. Neglecting for simplicity the second term in the brackets in (\ref{NT}), one may speculate that
\begin{equation}
    \tilde m^2 \lesssim \frac{ 10^{-11} \times 8 \pi^2}{g^2 \phi^2}
\end{equation}
in the model of~\cite{Nakayama:2011ri}. Assuming $g = 0.1$ and $\phi = 5 M_p$, one would find the upper bound $\tilde m \lesssim 6 \times 10^{-5}M_P$. Most phenomenological models of supersymmetry breaking would be consistent with this upper limit - it would even allow the ratio between the masses of the Starobinsky inflaton and its ``Starobinskino'' partner to be ${\cal O}(1)$.
However, the model studied in~\cite{Nakayama:2011ri} does not realize Starobinsky(-like) inflation.

 A no-scale supergravity model of Starobinsky(-like) inflation was proposed in~\cite{Ellis:2013xoa}. As has recently been emphasized~\cite{Antoniadis:2025pfa}, successful Starobinsky-like inflation requires a high degree of fine-tuning. The model proposed in~\cite{Ellis:2013xoa} is based on a Wess-Zumino superpotential with bilinear and trilinear terms, and requires a very specific value of the trilinear coupling~\cite{Ellis:2013xoa,Antoniadis:2025pfa}. This coupling is subject to wave-function renormalization, which is small in this model because the trilinear coupling is ${\cal O}(10^{-5})$, rendering the fine-tuning technically natural. However, the supergravity K\"ahler potential is also subject to radiative corrections, which require further investigation.

\section{Summary and Conclusions}
\label{summary}

We have considered in this paper three classes of one-loop radiative corrections to the Starobinsky inflationary potential. They include the Coleman-Weinberg-like correction generated by the inflaton itself and those due to possible couplings of the inflaton to fermions and bosons - recalling that one or the other is needed to reheat the Universe following inflation (or both). We have found that the Coleman-Weinberg-like correction is negligible during inflation: $\Delta V/V \sim H^2/M_P^2 \sim 10^{-10}$, and hence has no observable impact on the CMB observables $n_s$ and $r$.

However, the impact of inflaton decay couplings is potentially significant. In the case of a massless fermion, the {\it Planck} determination of $n_s$ imposes a 2$\sigma$ upper bound on the inflaton-fermion Yukawa coupling in the Einstein frame of $y< 4.5\times 10^{-4}$, corresponding to an upper bound on the reheating temperature of $T_{\text{RH}}< 2 \times 10^{11} \text{ GeV}$. On the other hand, the ACT determination of $n_s$ could be accommodated if the Yukawa coupling in the Einstein frame is constrained to be $3.8\times10^{-4}< y< 5.6\times10^{-4}$, corresponding to the reheating temperature constraint $1.7 \times 10^{11}<\trh< 2.8 \times 10^{11} \text{ GeV}$. We have also provided in Fig.~\ref{mfcontour} the {\it Planck} and ACT bounds on the possible fermion mass as functions of the fermion Yukawa coupling. For a coupling of order $y=10^{-4}$, the upper limit on the fermion mass is of order $10^{16}$~GeV. If one considers instead a fermion mass in the Jordan frame, it is exponentially suppressed at larger inflaton field values but the one-loop corrections may nevertheless have a significant impact on the CMB observables. We found the bounds $m_f \lesssim 5.8 \times 10^{16}$\ GeV from the {\it Planck} data and $5.4 \times 10^{16}\ {\rm GeV} \lesssim m_f \lesssim 6.2 \times 10^{16}$\ GeV from the ACT data.

The case of a massless scalar in the Einstein frame, coupled to the Starobinsky inflaton via a dimensionful trilinear coupling, is also interesting. The {\it Planck} data set an upper limit $\kappa \leq 4 \times 10^{12}$\ GeV, implying that the reheating temperature $\leq 4.2 \times 10^{13}$\ GeV, but the scalar loop corrections never push $n_s$ into the ACT range. In this case the radiative corrections cause the effective potential to decrease at large inflaton field values beyond a local maximum at a value $\phi_{max}$. This sets an upper limit on the initial value of the inflaton field (assuming it is initially stationary), which is larger than the pivot value $\phi_*$ over the range of $\kappa$ allowed by {\it Planck} data, as seen in Fig.~\ref{phimaxscalar}. For couplings of order $\kappa \sim 10^{12}$~GeV, we have also derived a limit of order $10^{16}$~GeV for a massive scalar coupled to the inflaton. Finally, we have also considered the limits if the inflaton coupling is generated by the conformal transformation from the Jordan frame to the Einstein frame. Indeed, the coupling to the Standard Model Higgs is generated this way leading to a reheating temperature of $2.9 \times 10^{9}$~GeV \cite{Ema:2024sit}. To ensure that the scalar coupled to the inflaton does not develop a VEV due to its conformal coupling to the inflaton, we require only $m_\chi \gtrsim m$. In order to remain consistent with the {\it Planck} 2018 determination of $n_s$, we further require $m_\chi \lesssim 6 \times 10^{16}$~GeV, effectively placing an upper limit to the GUT scale if GUT physics is introduced in the Jordan frame.

We have also discussed briefly the potential implications of supersymmetry for our analysis. In the limit of exact supersymmetry, the radiative corrections should vanish and there would be no constraints on the possible couplings of the inflaton to fermions and bosons. However, supersymmetry is broken at some scale $\tilde m$, leading to non-zero radiative corrections and the possibility that CMB data might set an upper limit on $\tilde m$. The Starobinsky model does not itself have a simple supersymmetric extension, though no-scale supergravity can accommodate avatars of the Starobinsky model \cite{Ellis:2013nxa,Ellis:2018zya}. We leave for future work a detailed consideration of radiative corrections in such models.

Our work shows that radiative corrections may have important effects on the predictions of the Starobinsky model, and potentially on other models of inflation. Radiative corrections merit more detailed consideration in future studies of inflationary models and their comparison with observational data.

\subsection*{Acknowledgments}
We thank Yohei Ema for discussions at the initial stages of this work.
The work of J.E. was supported by the United Kingdom STFC Grant ST/T000759/1. 
The work of T.G. and K.A.O. was supported in part by DOE grant DE-SC0011842 at the University of Minnesota. 
The work of K.K. was supported in part by Niigata University Grant for the Enhancement of International Collaborative Research, 2025.

\end{document}